\documentclass[doublecol]{epl2}

   \usepackage{epstopdf}

\usepackage{amssymb}
\usepackage{amsmath}
\usepackage{graphicx}

\makeatletter

\begin{document}
   \DeclareGraphicsExtensions{.pdf,.gif,.jpg.eps}

\title{Microscopic model for the semiconductor-to-ferromagnetic-metal transition in FeSi$_{1-x}$Ge$_{x}$ Alloys}
\author{
Kai-Yu Yang$^{1}$, Y. Yamashita$^{2}$, A.M. L\"auchli$^{3}$, M. Sigrist$^{1}$, and T. M. Rice$^{1}$
}

\institute{
\inst{1} Institut f$\ddot{u}$r Theoretische Physik, ETH Z$\ddot{u}$rich,CH-8093 Z$\ddot{u}$rich, Switzerland \\
\inst{2} College of Engineering, Nihon University, Koriyama 963-8642, Japan \\
\inst{3} Max Planck Institut f\"ur Physik komplexer Systeme, N\"othnitzer Str. 38, D-01187 Dresden, Germany
}


\abstract{
The simplified bandstructure introduced by Mazurenko et al to model FeSi is used to analyze the singlet semiconductor to ferromagnetic metal transition in the isoelectronic isostructural alloys, FeSi$_{1-x}$Ge$_x$. The complex bandstructure of the alloy is replaced by an alternating  chain of  doubly and singly degenerate atoms to represent Fe and Si/Ge respectively. The former(latter) form narrow(broad) bands with a substantial hybridization between them. A substantial onsite repulsion including a HundÕs rule coupling  is introduced on the Fe sites. The mean field phase diagram contains a first order phase transition from the singlet semiconductor to a ferromagnetic metal with increasing temperature and interaction strength similar to the alloys. The analysis also reproduces the rapid rise of the spin susceptibility in the semiconductor with a crossover to a Curie-Weiss form at higher temperatures. Good agreement is found at zero temperature between the mean field and  accurate DMRG
calculations.}

\date{\today}
\maketitle

\section{Introduction}
The rapid crossover around room temperature in FeSi from semiconducting to metallic behavior with strong magnetic fluctuations \cite{Jacarino-PR-67} has long been studied both experimentally \cite{Aeppli-95, Schlesinger-PRL-93,Mandrus-PRB-95, Paschea-PRB-97, Sluchanko-EPL-00, Yeo-PRL-03, Klein-PRL-08, Arita-PRB-08} and theoretically \cite{Takahashi-JPSJ-79, Evangelou-JPC-83, Mattheiss-PRB-93, Schlottmann-JAP-94, Anisimov-PRL-96, Fu-PRB-95, Takahashi-JPCM-97,Anisimov-PRL-02, Urasaki-PRB-98, Jarlborg-JMMM-04, Kunes-PRB-08, Mazurenko-PRB-10}. The small scale of the energy gap in the semiconductor has lead to proposals that this system is an unusual example of a Kondo insulator in a transition metal compound \cite{Aeppli-95, Mandrus-PRB-95, Schlottmann-JAP-94}. 
An alternative interpretation proposes that FeSi is a nearly ferromagnetic semiconductor \cite{Takahashi-JPSJ-79, Evangelou-JPC-83, Takahashi-JPCM-97}.
A variant on this proposal interpreted the anomalous behavior FeSi at room temperature and zero field as arising from proximity to the critical point of the first order transition from singlet semiconductor to ferromagnetic metal with increasing magnetic field. 
This proposal received support from a series of electronic structure calculations using the LDA+$U$
method~\cite{Anisimov-JPCM-97} which found that a ferromagnetic metallic state was close by in energy to the paramagnetic semiconducting
state~\cite{Anisimov-PRL-96}. Subsequently we realized that the isostructural isoelectronic compound
FeGe has a magnetic metallic groundstate \cite{Lundgern-PL-68, Lebech-JPCM-89} with a long period rotation of the magnetic moment due to a Dzyaloshinskii-Moriya interaction. The isostructural isoelectronic FeSi$_{1-x}$Ge$_x$ alloy system displays an anomalous
semiconductor to magnetic metal transition~\cite{Yeo-PRL-03, Anisimov-PRL-02}. This transition occurs  as the lattice is expanded. Note this behavior is opposite to the
usual Mott transition from a paramagnetic metal to a magnetic insulator with increasing lattice parameter. 

The earlier electronic structure calculations are restricted to zero temperature. A phenomenological extension 
to finite temperature was developed later in terms of interacting electrons and holes excited across the 
small band gap which gave a good description of the metal-insulator transition in the FeSi$_{1-x}$Ge$_x$ 
alloys \cite{Anisimov-PRL-02}. This left open however the question of the underlying microscopic model responsible for this unusual
behavior. It is not clear how to use the Kondo insulator proposal as basis for a microscopic model since
the band structure of FeSi shows strong, not weak, hybridization between the
Fe-3$d$ band and Si valence bands. Also the magnetism of the metal is essentially ferromagnetic and not 
antiferromagnetic, as would be expected from a RKKY-picture. For all these reasons it is desirable to 
have a microscopic model closely based on the band structure which can reproduce the observed
phase diagram.

\section{Simplified Multiband Model for F${\mbox{e}}$S${\mbox{i}}$}

%

Although FeSi is a cubic compound it has a complex unit cell containing 4 formula units with low point
symmetry at the Fe and Si sites. The {\it ab initio} LDA band structure calculations for FeSi correctly predict
a small gap semiconducting ground state~\cite{Anisimov-PRL-96,Mattheiss-PRB-93, Jarlborg-JMMM-04}.
This band structure disagrees with the Kondo insulator approach which supposes
a narrow Fe 3$d$ band weakly hybridizing with Si $s$, $p$ bands at the Fermi energy.
The semiconducting gap does not arise from weak hybridization between
the Fe-3$d$ and Si-3$p$ orbitals but lies with a complex of bands that are predominantly composed of 
non-bonding Fe-3$d$ orbitals. The whole band structure of FeSi is quite complex so that it is useful to have
a simpler model which contains the essential physics. To this end Mazurenko and collaborator \cite{Mazurenko-PRB-10} recently examined the Fe-Si hybridization in detail and then introduced a simplified model.
 Fe sites are represented by doubly degenerate orbitals and substantial 
onsite Coulomb correlations. These bands hybridize strongly with single orbital sites which form a broad band representing Si.
To keep matters as simple as possible we analyze a one dimensional  model with alternating Fe and
Si sites. As we will discuss below, this simplification is not essential. 
It has the advantage that it
allows us to use the powerful DMRG (density-matrix-renormalization-group) method \cite{DMRG-White} to check the
results that we obtain using a mean field approximation. 

\begin{figure}[tb]
\centerline
{
\includegraphics[width = 7.5cm, height =3.2cm, angle= 0] 
{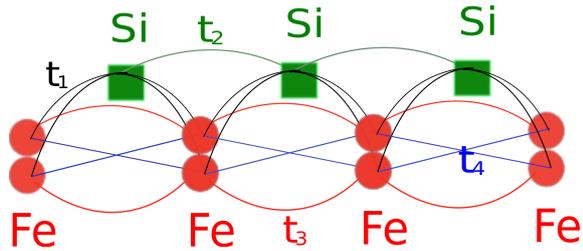}
}
 \caption[]
{(Color online) Sketch of the 1D FeSi model. The different line colors denote the 
various Fe-Fe and Fe-Si hopping processes. In the DMRG calculations 
we adopt open boundary conditions and terminate the chain by putting 
a Si orbital at each end.}
\label{fig:lattice}
\end{figure}

The simple multiband model is illustrated in Fig.~\ref{fig:lattice}.
The Fe sites are represented by doubly
red circles and Si sites by single dark green square. A narrow band width is chosen for the Fe-sites,
a broad band width for Si sites and a substantial Fe-Si hybridization is introduced. A substantial Coulomb
interaction with onsite Hund's rule exchange is included on Fe sites while interactions on Si sites are 
ignored. To keep matters  as simple as possible we consider only a one dimensional model with an alternating
chain of Fe and Si sites. Since we are interested in the competition between ferromagnetism
and paramagnetism special features of one dimension such as perfect nesting are not relevant. 
In the mean field theory presented here we restricted our attention to solutions without orbital polarization, i.e $\langle n_{m,\sigma} \rangle$ on Fe is always independent of the orbital index $m$. Actually we found a lower energy within mean field theory could be achieved by relaxing this condition, e.g. allowing states without net spin and orbital polarizations through compensating changes in the spin and orbital density such that $n_{m, \sigma} =  n_{\bar{m}, -{\sigma}} $$ \ne $
$n_{ m, -\sigma} =  n_{ \bar{m}, {\sigma}} $ on Fe sites. These states build in onsite spin-orbital correlations. We argue these states should not be compared to the simple mean field states we consider here. This conclusion is supported by the good agreement between the form of the mean field and exact DMRG phase diagrams.

The Hamiltonian consists of kinetic energy ($H_0$) and interaction ($H_1$) terms. The former is
parametrized by the hopping elements $ t_1, \dots , t_4 $ illustrated in Fig.~\ref{fig:lattice}. 
The resulting band energies $ \varepsilon_i (k) $ are obtained by diagonalizing a $3\times3 $-matrix, 
$ \varepsilon_k^{mm'} $, where $ m=1,2 $ and $ m=3 $ denote Fe $d$- and Si $p$-orbitals, respectively. 
\begin{equation}
  \varepsilon_k^{mm'} = \left( \begin{array}{lll} 
      2 t_3 \cos k & 2 t_4 \cos k & 2 t_1 \cos \frac{k}{2} \\
      2 t_4 \cos k & 2 t_3 \cos k & 2 t_1 \cos \frac{k}{2} \\
      2 t_1 \cos \frac{k}{2} & 2 t_1 \cos \frac{k}{2} & 2 t_2 \cos k 
    \end{array} \right)
  \label{Eq-1}
\end{equation}
The hopping parameters from the band structure are obtained after the projection procedure \cite{Mazurenko-PRB-10}:
$t_{1}$=2.0 eV, $t_{2}$=3.0 eV, $t_{3}$=-1.2 eV, $t_{4}$=-0.25 eV. In the following context, without special claim, all energy variables are in units of eV.

\begin{figure}
\vspace{0.5cm}
\centerline{\includegraphics[width = 5.5cm, height =8.5cm, angle= 270] {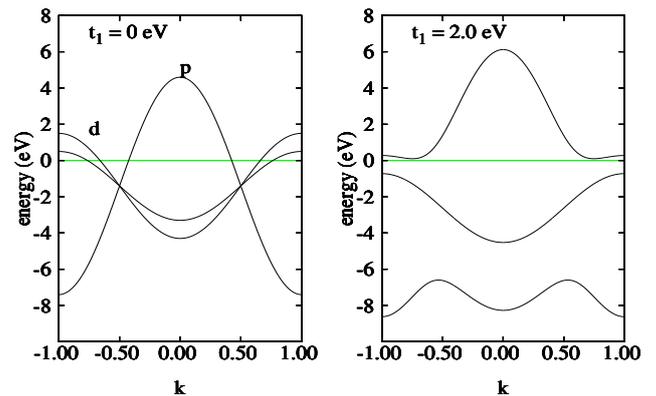}}
\caption{ (Color online) Band structure of the simplified FeSi model for the case without Fe-Si
hybridization (left panel) and with strong Fe-Si hybridization (right panel).}
\label{fig:dispersion}
\end{figure}
The resulting energy bands are illustrated in Fig.~\ref{fig:dispersion}. In the left panel the bands appear without 
hybridization. In the right panel the effect of a substantial Fe-Si hybridization is to split the bands into a lower
bonding and an upper anti-bonding bands with a non-bonding band in between. When we choose a filling
of 4 electrons per unit cell, the Fermi energy lies in the small energy gap between the non-bonding
and anti-bonding bands. Thus, we obtain a small gap semiconducting ground state 
when interactions are ignored. Note the small value of the energy gap is not a consequence of a weak
Fe-Si hybridization. Rather it separates two Bloch states with predominantly Fe-character arising from the two orbitals on the Fe-sites.

One consequence of the choice of a one dimensional model is the divergence of the density of
states (DOS) at the band edges. We believe this is not important 
for two reasons. First, as will be discussed below, the transition between the semiconducting and the ferromagnetic
metallic phases is a first order transition to a ferromagnet with a substantial moment
so the exact form of the DOS peak is not relevant. Secondly the full FeSi band structure calculations
also show narrow peaks in the DOS at the band edges in the semiconducting phase. 

\section{Mean field approximation at T=0}
The interaction term in the Hamiltonian is introduced only on Fe sites. It takes the standard multi-orbital Hubbard form 
including both intra- and inter-orbitals onsite Coulombs interactions
\begin{equation} \begin{array}{ll}
    H_I &=  \displaystyle 
      U \sum_i (  n_{i 1 \uparrow} n_{i1 \downarrow} + n_{i2 \uparrow} n_{i2 \downarrow} ) \\
    & \displaystyle
    + U' \sum_i (n_{i1 \uparrow} + n_{i1\downarrow})(n_{i2 \uparrow} + n_{i2\downarrow}) \\
    & \displaystyle
    - J \sum_i (n_{i1 \uparrow} n_{i2 \uparrow} + n_{i1 \downarrow} n_{i2 \downarrow}) \\
    & \displaystyle
    + J \sum_{i,m=1,2} ( c_{im \uparrow}^{\dag} c_{i \bar{m} \uparrow} c_{i \bar{m} \downarrow}^{\dag} c_{i m \downarrow} + c_{im \uparrow}^{\dag} c_{i \bar{m} \uparrow} c_{i m \downarrow}^{\dag} c_{i \bar{m} \downarrow}  )
  \end{array}
  \label{eqn:Interactions}
\end{equation}
The intra- and inter-orbital onsite Coulomb interactions $U$ and $U'$ are related through 
the Hund's rule exchange energy $J$ by $ U=U'+2J $. As we shall discuss below the Hund's rule exchange energy, J, plays an important role in stabilizing ferromagnetism. Several earlier studies which focussed on the anomalous T-dependence of paramagnetism in FeSi omitted the inter-orbital Hund's rule coupling \cite{Fu-PRB-95, Urasaki-PRB-98, Kunes-PRB-08, Mazurenko-PRB-10}. 

\begin{figure}
  \centerline{\includegraphics[width = 5.5 cm, height = 10.5cm, angle = 270 ]{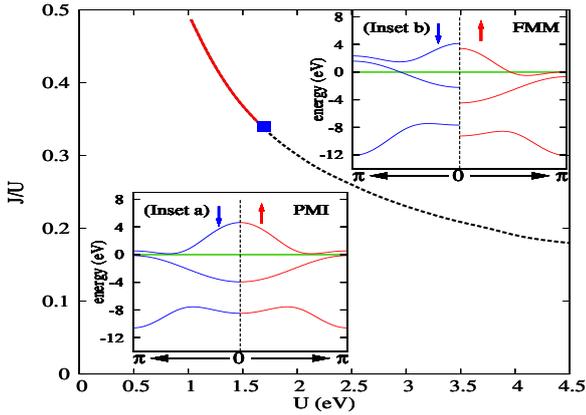}}
  \caption{(Color online) Zero-temperature mean field phase diagram for $0<U<5$eV. The dashed line (black) is the first order and the solid line (red) the second order transition boundary between the  paramagnetic insulating state (PMI) and the ferromagnetic metallic state (FMM). They are separated by the tricritical point (square blue dot). 
Insets: energy bands of the chain (a) PMI with $U =$ 2eV, $J =$ 0.5eV and (b) FMM with $U =$ 4eV, $J =$ 1eV. 
   \label{fig:MeanField1}
  }
\end{figure}
\begin{figure}
  \centerline{\includegraphics[width = 5.5 cm, height = 10.5cm, angle = 270 ]{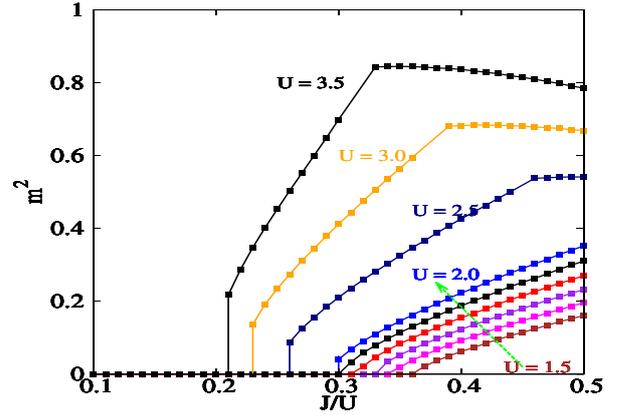}}
  \caption{(Color online) Square of the ferromagnetic moment, $m^{2}$, versus J/U at zero temperature. The tricritical point with $J/U \sim 0.3$ and $U_{c} \sim 1.7$eV (blue square dot in Fig.\ref{fig:MeanField1}) marks the transition from a first order transition at large U (small J/U) to a second order transition at small U (large J/U).
   \label{fig:SZT0}
  }
\end{figure}
\begin{figure}
  \centerline{\includegraphics[width = 5.5 cm, height = 10.5cm, angle = 270 ]{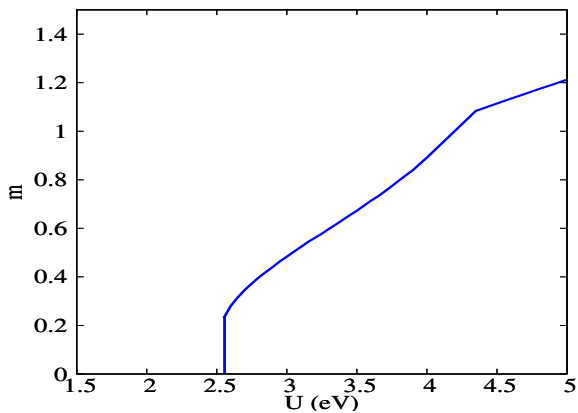}}
  \caption{(Color online) Ferromagnetic moment $m$ versus $U$ at constant $J/U$ = 0.25 and $T=0$.
   \label{fig:SZT0J025}
  }
\end{figure}

First, we analyze the combined Hamiltonian $ H_0 + H_I $ at $ T=0 $ using a mean field
approximation for the case of 4 electrons per unit cell. The paramagnetic small gap semiconducting (PMI)
phase remains stable initially with a reduced energy gap as the interactions are introduced. In this phase as apparent 
in the left panel of Fig.~\ref{fig:MeanField1} the 4 electrons are distributed between Fe and Si sites with density
$ n_{Fe} : n_{Si} = 3 : 1 $. Upon increasing the interaction strength there is a transition 
to a ferromagnetic metallic phase (FMM) with a moment primarily on the Fe sites, $ m_{Fe} = \sum_m \langle n_{Fe, m \uparrow} - n_{Fe, m \downarrow} \rangle $. This leads to a splitting of the bands between minority and majority spin directions as illustrated in right inset of Fig.~\ref{fig:MeanField1}. The phase boundary in Fig.~\ref{fig:MeanField1} between these phases depends on the ratio
$ J/U $ which we denote as $ \alpha $. 
Along the phase boundary between PMI and FMM the transition is first order at large $U$ and small $\alpha$ but turns into a second order transition for values of $ \alpha $ greater than a critical value, $ \alpha_c \approx 0.3 $ as shown in Fig.~\ref{fig:SZT0}.
 The dimensionless parameter $\alpha =J/U $ should remain almost constant passing from FeSi to FeGe. The value of $U$ is to be compared to the bandwidth, therefore increases in FeGe relative to FeSi due to the increase in the lattice constant. Thus, the change between two materials would follow along a horizontal line in Fig.~\ref{fig:MeanField1}. 
 The bands in Fig.~\ref{fig:MeanField1} inserts, illustrate the change between a PMI and FMM, at PMI side in the left panel with a reduced energy gap for triplet excitations and the FMM side in the right inset panel. 
  The transition in the magnetic momentum at a constant $\alpha = 0.25$ and increasing $U$ is illustrated in Fig.~\ref{fig:SZT0J025}. Initially there is a jump followed by a rapid rise to a value of $m_{A} \simeq 1$ which agrees with the observed moment of $\approx 1 \mu_{B} $ per Fe in the magnetic phase of FeGe~\cite{Yeo-PRL-03,Lundgern-PL-68,Lebech-JPCM-89}.
This unusual value is a consequence of the strong Fe-Si hybridization which causes only a partial polarization of a single band. Further increases in $U \ge 4.1$eV lead to a slower increase in the polarization since the Fermi energy in the majority spin bands enters a region with a rapidly rising dispersion. Note the earlier LDA+U calculations \cite{Anisimov-PRL-02, Anisimov-PRL-96} with the full FeSi band structure gave a similar moment $\sim $ 1$\mu_{B}$.

If the interaction strength is strongly increased a new antiferromagnetic phase is stabilized ($U >5.9$eV). In this 
parameter region the strong onsite Coulomb repulsion on the Fe sites causes an electron to be
transferred to the Si site resulting in the distribution $ n_{Fe} : n_{Si} = 2 : 2 $. The spins of the two 
electrons on the Fe site are aligned parallel according to Hund's rule. Superexchange through the Si
sites aligns the spins on successive Fe sites antiparallel resulting in an antiferromagnetically ordered 
insulator. However, this phase which appears only at large values of $U$, is not relevant to FeSi or FeGe.
Note, the relevant values of the onsite interaction, U, here are larger than the value of 1eV used by Mazurenko et al [21] in their Dynamic Mean Field Theory calculations and fits to the optical properties and spin susceptibility of FeSi.

\section{Finite Temperature Results}
\begin{figure}
  \centerline{\includegraphics[width=0.6\linewidth, angle = 270]{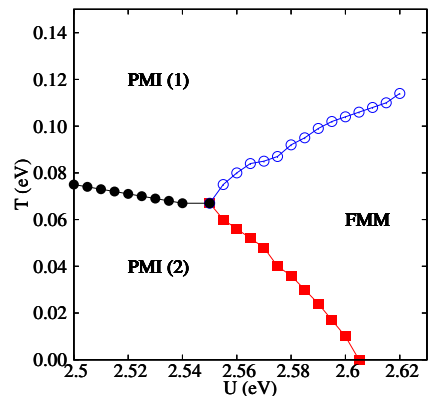}}
  \caption{  (Color online)  Mean field phase diagram in (T,U) space with $J/U = 0.25$ with $U_{c} \sim 2.6$eV and a tricritical point at $U_{t} \sim 2.55$eV. The square dots (red) mark the first order transition between paramagnetic insulating and ferromagnetic metallic phase (PMI and FMM), and the empty circles (blue) stand for the second order transition. The filled circles (black) indicate the temperature of the maxima in the spin susceptibility $ \chi(T) $.
   \label{fig:MeanField-TU}
  }
\end{figure}
\begin{figure}
  \centerline{\includegraphics[width=0.6\linewidth, angle = 270]{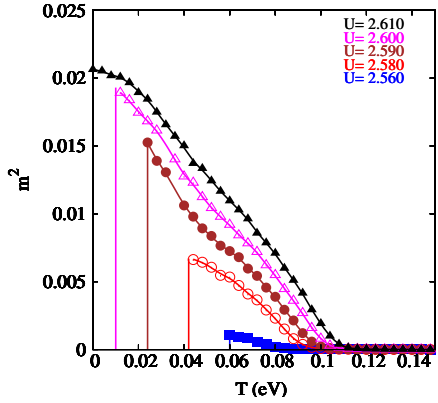}}
  \caption{
    (Color online)  Square of the ferromagnetization, $m^{2}$ as function of temperature for $U \gtrsim U_{t} \sim 2.55 $ eV with $J/U = 0.25$.
    \label{fig:sz}
  }
\end{figure}

We turn now to the temperature effects within the mean field approximation. The earlier model for FeSi based on LDA+U calculation \cite{Anisimov-PRL-96}, placed FeSi in the PMI phase with the FMM phase nearby at a slightly larger value for $U$. Extending the mean field results to finite T for fixed $\alpha = 0.25$ and values of $U$ near the critical value $U_{c} = 2.6$eV, leads to the (T,U) phase diagram shown in Fig.~\ref{fig:MeanField-TU}. At small values of T and $U \le U_{c}(T=0)$ there is a first order transition from the PMI phase to the FMM phase at finite T driven by the larger entropy of the FMM phase. At higher temperatures the ferromagnetic order disappears in a second order transition to a paramagnetic insulating phase. These two transitions are illustrated in Fig.~\ref{fig:sz} which shows the temperature dependence of the ferromagnetic moment $m_{Fe}(T)$ at fixed $\alpha$ and $U$. As $U$ is decreased further from $U_{c}(T=0)$, the temperatures of the first and second order transitions approach each other and come together at a tricritical point located at $T_{t}= 0.07$eV, and $U_{t} = 2.55$eV.

\begin{figure}
  \centerline{\includegraphics[width=0.6\linewidth, angle = 270]{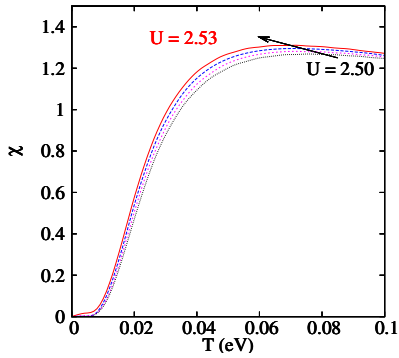}}
  \caption{
  (Color online) Spin susceptibility $\chi(T)$ for $U \lesssim U_{t} $ with $J/U = 0.25$, the four curves are for $U=2.50, 2.51, 2.52, 2.53$eV. The maxima of $\chi(T)$ correspond to the crossing points between the curves and the arrow.
    \label{fig:chi}
  }
\end{figure}

For small values of $U$ ($<2.55$eV), the  results of calculation of the magnetic susceptibility, $\chi(T)$ are shown in Fig.~\ref{fig:chi}. The locus of the susceptibility maxima, $T_{max}(U)$  are plotted in Fig.~\ref{fig:MeanField-TU}. We see that $T_{max}(U) $ extrapolates to the tricritical point. This behavior is consistent with explanation~\cite{Anisimov-PRL-96} of the maximum in $\chi(T)$ observed in FeSi at 500K by Jacarino et al,~\cite{Jacarino-PR-67}, as arising from the proximity to a critical end point of a line of first order transition between PMI and FMM phase in a magnetic field.

\section{Density Matrix Renormalization Group simulations}


\label{dmrg}
In order to test the mean field analysis  we used a numerically 
exact simulation of the one-dimensional microscopic model for FeSi defined by Eqs.~(\ref{Eq-1})
and (\ref{eqn:Interactions}), based on the density matrix renormalization group (DMRG) algorithm~\cite{DMRG-White}.  

{\em DMRG Setup ---}
The geometry used in the computations is displayed in Fig.~\ref{fig:lattice}.
It consists of one Fe atom (containing two orbitals) and one Si atom per unit cell and we use a setup where the chain is
terminated by a Si atom at both ends.

While we are interested in the case of an average density of 4 electrons per unit cell, one has to be careful in
choosing the correct filling for a finite system with open boundary conditions, in order for the lowest two bands
to be filled. For the chosen setup the appropriate number of electrons is $N_e=4N_{c}+2$, where $N_{c}$ 
denotes the number of unit cells.

\begin{figure}
  \centerline{\includegraphics[width=0.9\linewidth]{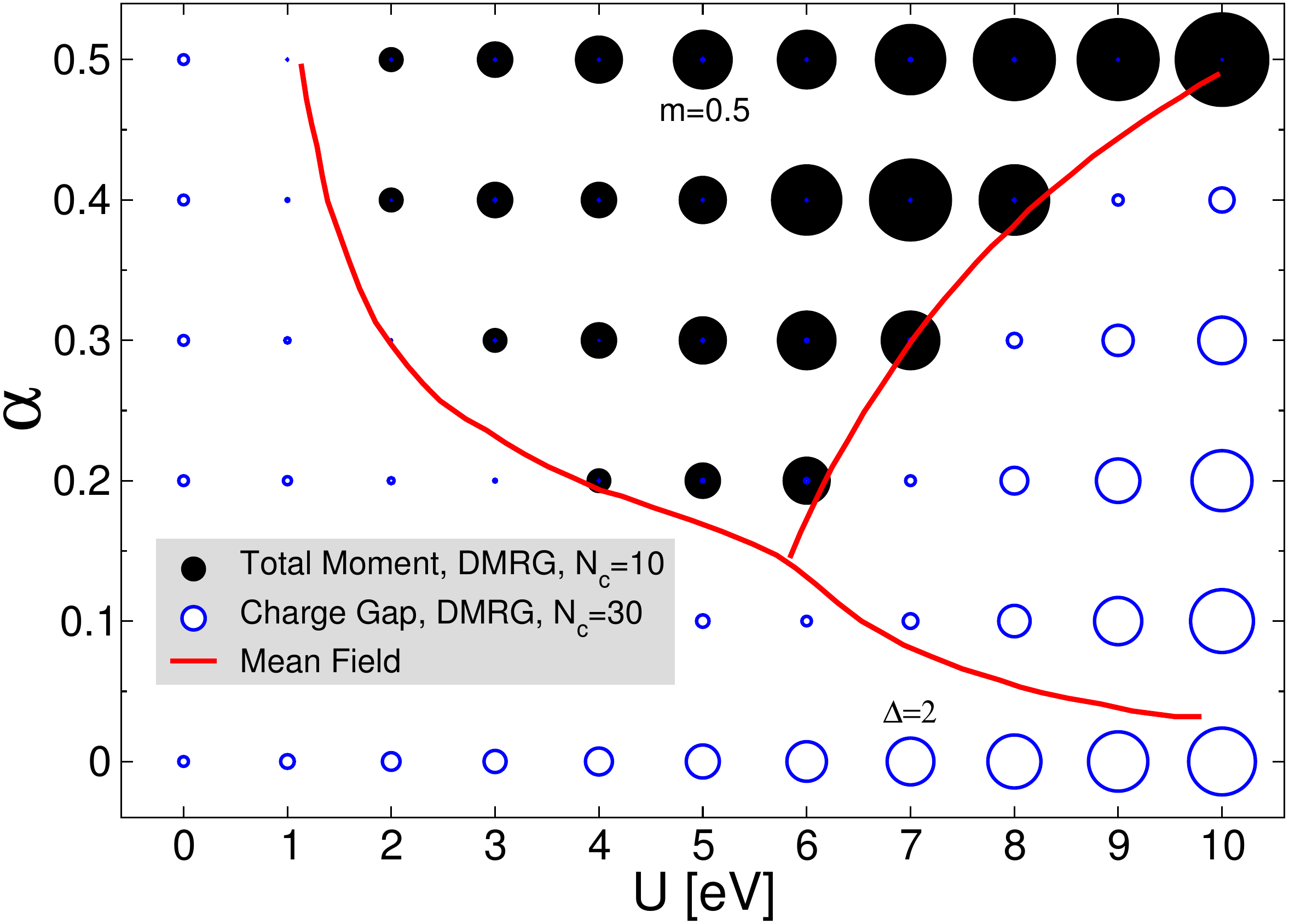}}
  \caption{
  \label{fig:SpinPhaseDiag} Phase diagram obtained from density matrix renormalization 
  group simulations. Filled circles denote the size of the partially polarized ferromagnetic moment.
  Empty circles indicate the size of the charge gap. Two of circles have been labelled with the value scaling to
  their diameter ($ m=0.5 $ for the black circles and $ \Delta = 2 $ for empty circles). The solid  lines show the mean field phase boundaries which are extended to higher values of $ U $ and include the antiferromagnetic state at $ U > 5.9 eV $.
  }
\end{figure}

Our main interest is to detect and locate the partially polarized conducting state reached
upon increasing the correlations on the Fe sites. The mean-field results 
predict such a phase for intermediate values of $U$ and $\alpha$ larger than $0.2$. The 
polarized state can be found in DMRG by scanning the energy in each $S^z$ sector. The
degeneracy of the energy with respect to $S^z$ gives the spin multiplet $S$ of the groundstate.
In Fig.~\ref{fig:SpinPhaseDiag} we present the numerical phase diagram for a system consisting 
of $N_{c}=10$ unit cells (21 atoms), where the diameter of the filled circle is proportional to the
size of the magnetic moment. In addition we have measured the charge gap
$\Delta_N=\frac{1}{2}[E(N_e+1)+E(N_e-1)]-E(N_e)$, which we take as a measure for insulating ($\Delta_N>0$)
or conducting behavior ($\Delta_N=0$). The charge gap is shown for $N_{c}=30$ in Fig.~\ref{fig:SpinPhaseDiag} 
using empty circles. While the charge gap at $U=0$ corresponds to the value of the small-gap semiconductor,
the gap seems to close in the partially polarized region as expected, and then opens again and reaches
large values for large $U$. The agreement of the DMRG-based boundary of the partially polarized phase
with the mean field results is remarkable.

The mean field theory predicts also an antiferromagnetic insulating phase in addition to the semiconducting and
the partially polarized phases. In the one-dimensional model considered here, a long-range antiferromagnetically ordered phase 
is not present, but is replaced by a Haldane-like phase. We have checked for a specific
value of the interactions ($U=10$, $\alpha=0.2$) that the system has a finite spin gap, edge-localized
spin excitations, and a two-fold degenerate entanglement spectrum (in the most probable charge sector)~\cite{HaldaneES}, as
expected for a Haldane phase. The region of the phase diagram in the vicinity of the semiconductor ($U=0$) has also a spin gap,
however no edge spin excitations have been detected and the entanglement spectrum does not exhibit a clear two-fold
degeneracy. It will be an interesting question to explore in the future whether the two phases are adiabatically connected or separated by
a phase transition. Both possibilities can in principle occur, as suggested in Ref~\cite{AnfusoHaldane}.

In conclusion we find that the singlet semiconductor to ferromagnetic transition observe in FeSi$_{1-x}$Ge$_x$ alloys can be well reproduced by a simplified one dimensional model of the bandstructure which retains the strong inter-atom  hybridization of the full bandstructure. The HundÕs rule term in the onsite interaction on Fe sites plays an important role in stabilizing the ferromagnetic state. The spin susceptibility of the paramagnetic semiconductor rises strongly at low temperature with a maximum at a temperature which connects on to the tricritical point of the Curie temperature of the ferromagnetic metal as the onsite interaction is increased. The reliability of the mean field approximation was tested against accurate DMRG at zero temperature with good results.

\textbf{Acknowledgements}
We are grateful to V. Anisimov and G. Mazurenko for important discussions and input in the early stages of this work. We also acknowledge financial support from the Swiss Nationalfonds and the NCCR MaNEP.


\begin{thebibliography}{99}

\bibitem{Jacarino-PR-67}  
Jaccarino V., Wertheim G.K., Wernick J.H., Walker L.R., and Arajz S., 
Phys. Rev. \textbf{160} (1967), 476.

\bibitem{Aeppli-95} Aeppli G.  and Fisk Z. , Comments Condens. Matter Phys. \textbf{16} (1992), 155.

\bibitem{Schlesinger-PRL-93} Schlesinger Z. et al., Phys. Rev. Lett, \textbf{71} (1993), 1748.

\bibitem{Mandrus-PRB-95}  
Mandrus D., ~Sarrao J.L., ~Migliori A., ~Thompson J.D., and ~Fisk Z., 
Phys. Rev. B \textbf{51} (1995), 4763.

\bibitem{Paschea-PRB-97}Paschea  S. et al, Phys. Rev. B, \textbf{56}, (1997), 12916.

\bibitem{Sluchanko-EPL-00} Sluchanki N. E.  et al, Euro Phys. Lett. \textbf{51} (2000), 557.

\bibitem{Yeo-PRL-03} Yeo S.  {\it et. al.}, Phys. Rev. Lett. \textbf{91} (2003), 046401.

\bibitem{Klein-PRL-08} Klein M. , et al, Phys. Rev. Lett. \textbf{101} (2008), 046406.

\bibitem{Arita-PRB-08} Arita M.  et al., Phys. Rev. B \textbf{77} (2008), 205117.

\bibitem{Takahashi-JPSJ-79} Takahashi Y.  and  Moriya T., J. Phys. Soc. Jpn. \textbf{46} (1979), 1451. 

\bibitem{Evangelou-JPC-83} Evangelou S.N.  and Edwards D.M. , J. Phys. C \textbf{16} (1983), 2121.

\bibitem{Mattheiss-PRB-93} Mattheiss L.F. and ~Hamann D.R., Phys. Rev. B \textbf{47} (1993), 13114.

\bibitem{Schlottmann-JAP-94} Schlottmann P. , J. Appl. Phys. \textbf{75} (1994), 7044.

\bibitem{Anisimov-PRL-96}
Anisimov V.I., ~Ezhov S.Y., ~Elfimov I.S.,~Solovyev  I.V., and ~Rice T.M., 
Phys. Rev. Lett. \textbf{76} (1996), 1735.

\bibitem{Fu-PRB-95} Fu C., and Doniah S., Phys. Rev. B \textbf{51} (1995), 17439.

\bibitem{Takahashi-JPCM-97} Takahashi Y. , J. Phys: condens. Matter \textbf{9} (1997), 2593.

\bibitem{Urasaki-PRB-98} Urasaki  K. and  Saso T., Phys. Rev. B \textbf{58} (1998), 15528.

\bibitem{Anisimov-PRL-02}
Anisimov V.I., ~Hlubina R., ~Korotin M.A., ~Mazurenko V.V., 
~Rice T.M., ~Shorikov A.O., and ~Sigrist M., 
Phys. Rev. Lett. \textbf{89} (2002), 257203.

\bibitem{Jarlborg-JMMM-04} Jarlborg T. , J. Mag. Mag. Mat. \textbf{283} (2004), 238.

\bibitem{Kunes-PRB-08}Kunes  J.  and Anisimov  V. T., Phys. Rev. B \textbf{78} (2008), 033109.

\bibitem{Mazurenko-PRB-10} Mazurenko  V. V. et al., Phys. Rev. B \textbf{81} (2010), 125131.

\bibitem{Anisimov-JPCM-97}
Anisimov V.I., ~Aryasetiawan F., and ~Lichtenstein A.I.,
J. Phys.: Condens. Matter \textbf{9} (1997), 767.

\bibitem{Lundgern-PL-68} Lundgern L., Blum  K.A. and Beckman O. B., Phys. Lett. \textbf{28A} (1968), 175.

\bibitem{Lebech-JPCM-89} Lebech B., Bernard J. and Frelhoft T., J. Phys. Condens. Matter \textbf{1} (1989), 6105.

\bibitem{DMRG-White}
White S. R., Phys. Rev. Lett. \textbf{69} (1992), 2863; 
Phys. Rev. B \textbf{48} (1993), 10345;
Schollw\"ock, U. Rev. Mod. Phys. \textbf{77} (2005), 259.

\bibitem{HaldaneES}
Pollmann F., Berg E., Turner A.M., and Oshikawa M.,
Phys. Rev. B \textbf{81} (2010), 064439.

\bibitem{AnfusoHaldane}
Anfuso F. and Rosch A.,
Phys. Rev. B \textbf{75} (2007), 144420.

\end{thebibliography}
\end{document}